\documentclass[aps,prb,twocolumn,showpacs,superscriptaddress]{revtex4}
\usepackage[english]{babel}
\usepackage{amssymb}
\usepackage{amsmath}
\usepackage{graphicx}
\usepackage{graphicx}
\usepackage{color}
\usepackage{dcolumn}
\usepackage{bm}

\begin{document}

\title{Mechanical lattice instability and thermodynamical properties in classical solids}

\author{G. Rastelli}
\affiliation{Univ. Grenoble 1/CNRS, LPMMC UMR 5493, Maison des Magist\`{e}res, 38042 Grenoble, France}

\author{E. Cappelluti}
\affiliation{Instituto de Ciencia de Materiales de Madrid (ICMM), CSIC,
c. Sor Juana In\'es de la Cruz 3,  Cantoblanco, E-28049 Madrid, Spain}
\affiliation{Istituto dei Sistemi Complessi (ISC), CNR,
v. dei Taurini 19,  00185 Rome, Italy}

\begin{abstract}
In this paper we revisit the onset of the instability of the solid state 
in classical systems within self-consistent phonon theory (SCPT). 
Spanning the whole phase diagram versus volume and versus pressure, 
we identify two different kinds of mechanism: 
one mainly relevant at constant volume, associated with the vanishing 
of the SCPT solution; and one related to the disappearing at a spinodal temperature 
of the solid phase as a metastable energy minimum. 
We show how the first mechanism occurs at extremely high temperatures and it is 
not reflected in any singular behavior of the thermodynamical properties. 
In contrast,  the second one appears at physical temperatures  which correlate well with 
the melting temperature and it is signalized by the divergence of the
thermal compressibility as well as of the the lattice expansion coefficient.
\end{abstract}


\date{\today}

\maketitle

\section{Introduction}

\begin{figure}[b]
\includegraphics[scale=0.25,angle=0.]{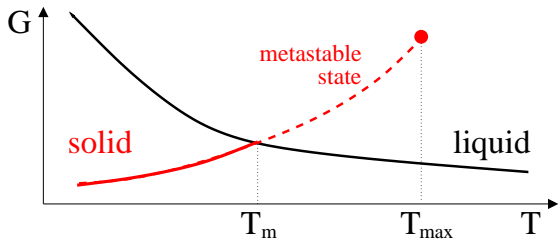}
\caption{(color online)  Pictorial evolution of solid/liquid
Gibbs free energy $G$ as a function of temperature $T$. At the melting
temperature $T_m$, the Gibbs free energies of the solid and liquid phases 
are equal, the system undergoes a first order transition.
A metastable solid phase (dashed line) can be still obtained 
up to a maximal point $T_{\rm max}$.}
\label{fig:1}
\end{figure}

The solid-liquid transition is one of the most common and extensively 
studied phase transition in condensed matter.
As a first order transition, overheating the solid  beyond 
the melting temperature is possible in 
nature,\cite{Daeges:1986,Grabaek:1992,Zhang:1998,Zhang:2000,Luo-1:2003,Sokolowski:2003,Siwick:2003,Iglev:2006,Mei:2007,Feng:2008} 
following  the metastable equilibrium state (Fig. \ref{fig:1}).

This route is however limited by the intrinsic instability of the
metastable phase one is considering.
Focusing on the metastable solid phase, a priori the maximum temperature of 
overheating (for a surface-free and perfect crystal without any defect or impurity)
could be identified with the  temperature 
above which the solid phase is not a sustainable phase.
The breakdown of the solid phase independently of the competition 
with the liquid phase is thus associated with the concept of 
intrinsic instability of the solid phase, i.e. with the breakdown of the 
conditions which make a solid phase sustainable even at a metastable level. 

Historically,  the problem of the {\em mechanical melting}\cite{Cahn:1988-89-2001} of a solid was 
discussed indirectly by the one-phase theories addressing the standard  
melting.\cite{Dash:1999,Lowen:1994,Lindemann:1910,Born:1939,Fecht:1988,Tallon:1989,Boyer:1985}  
The most two famous theories are  the well-known Lindemann criterion\cite{Lindemann:1910} and 
the Born rigidity catastrophe.\cite{Born:1939} 
Although these theories cannot explain the standard melting, 
they have been reconsidered, with a modern point of view, 
for a theoretical description of the instability of an overheated solid. 
Such a phenomenon, which was once thought to be unobtainable, has become now practical 
due novel experimental advances in heating techniques and in the fabrication 
of special samples, as it has been reported in a wide number of different 
systems.\cite{Daeges:1986,Grabaek:1992,Zhang:1998,Zhang:2000,Luo-1:2003,Sokolowski:2003,Siwick:2003,Iglev:2006,Mei:2007,Feng:2008} 
These experiments have stimulated a renewed theoretical  interest 
for the problem of the {\em mechanical melting}.

Recent molecular dynamics simulations have been also carried out to clarify the underlying 
microscopic mechanism which sets the stability limits of a overheated solid.\cite{Jin:2001,Luo-2:2003,Forsblom:2005,Luo:2007}  
In particular, Jin and coworkers found that above the equilibrium melting point $T_m$, 
local lattice instabilities governed by both Born and Lindemann criteria occurs at 
a well defined mechanical melting temperature $T_{\rm max}$,\cite{Jin:2001} 
at which the material cannot  survive in crystalline  order for any finite time interval.
Interestingly, the temperature of the mechanical melting was also
associated with a sudden and drastic rise of the atomic volume corresponding
to a large peak of the 
compressibility of the system.\cite{Jin:2001}

The crystal instability comes ultimately from anharmonic effects that may 
soften the lattice when the thermal fluctuations are large.
To this issue, the mechanical melting was also addressed in the past by using extensively 
the self consistent phonon theory 
(SCPT),\cite{Choquard:1967,Gillis-1:1968,Gillis-2:1968,Kugler:1969,Klein:1970,Gillis:1972,
Matsubara:1977,Pietronero:1979,Hasegawa:1980,Jayanthi-1:1985,Jayanthi-2:1985,Gong:1987,Moleko:1983}
which represents a suitable method to take into account the anharmonic effects
of the atomic oscillations.
Within this microscopic approach, the breaking of the self-consistent solution of the SCPT
is thought to represent the mechanical instability of the solid phase and
it defines a maximum temperature $T^*$  which was commonly assumed as upper estimation of 
the melting temperature (the variational free energy obtained by the SCPT is an upper bound  of 
the exact free energy).
However the values of the temperatures $T^*$ obtained in this way are 
much higher than the experimental  melting temperatures $T_m$
(even of two orders of magnitude) observed 
in several systems.\cite{Moleko:1983}
Moreover, they compare badly also with the experimental overheating temperatures $T_{\rm max}$ 
which are generally located between $T_m$ and $\sim 1.5 \, T_m$.\cite{Luo-2:2003}

\begin{figure}[t]
\includegraphics[scale=0.25]{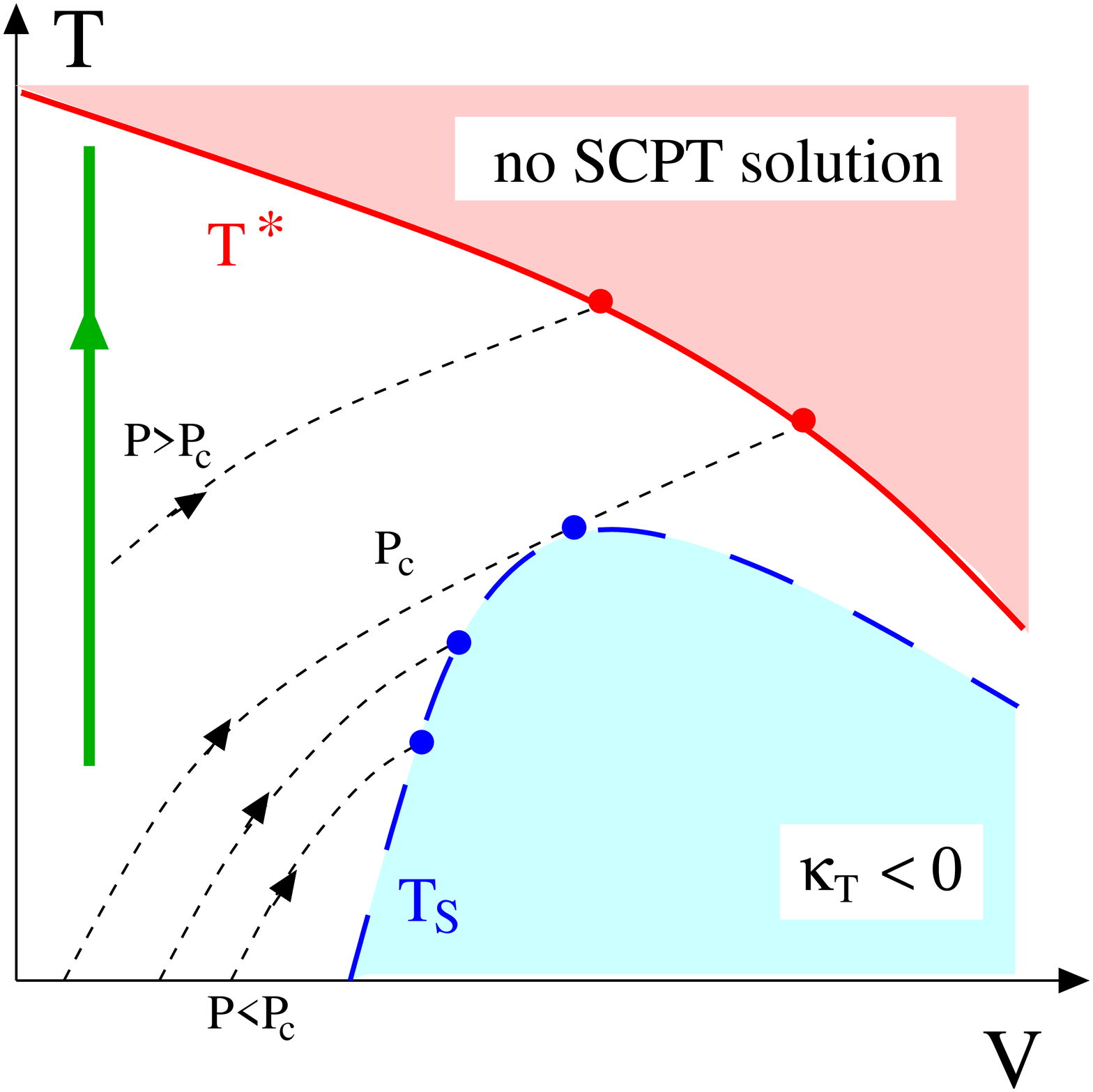}
\caption{(colors online) 
Schematic phase diagram for the breakdown of the mechanical stability 
of the solid phase  within the SCPT. Increasing the temperature at fixed volume 
(solid green vertical  line), the solid phase disappears at $T^*$ which
represents, in this case, the highest temperature at which the variational 
approach (SCPT) has a self-consistent solution (red line). 
Due to the lattice expansion, the mechanical instability at fixed (reasonable) 
pressure ($P \le P_c$) is first encountered in terms of the disappearing of the metastable
solid state, defining a spinodal temperature $T=T_S(P)$  (blue long dashed 
line) within a  well-defined SCPT solution. 
Below the line $T_S(P)$ the compressibility $\kappa_T$ is negative.}
\label{f-phd}
\end{figure}

In this paper we revise the self-consistent phonon theory focusing on
its extension for crystals at constant pressure $P$, 
taking into account in particular the lattice volume variation
as a function of both temperature and pressure.
Our main results are summarized in the schematic phase diagram
depicted in Fig. \ref{f-phd}.

In particular we find that within the SCPT  the physical description of the mechanical 
instability  of the solid phase approached at constant volume is qualitative and
quantitative different from the mechanical instability achieved at constant pressure.
In the first case, the instability of the solid phase occurs at rather high temperatures and 
it is associated with a breakdown of the SCPT solution at $T^*$, not reflected in any singular 
behavior of the thermodynamical quantities.
On the other hand, the mechanical instability approached at fixed pressure
can be properly interpreted in terms of a spinodal temperature 
$T_S$ at which the solid phase, always defined, disappears
as a metastable minimum.
In the latter case the mechanical instability is thus reflected in the divergence of 
the thermal compressibility $\kappa_T$ as well as in the divergence
of the lattice expansion coefficient $\alpha$ at $T=T_S$. 
This means that small fluctuations of the two  controlling fields $(T,P)$ produce 
large volume fluctuations,  $\Delta V/V = \alpha \Delta T$ and similarly  
$\Delta V/V = -\kappa_T \Delta P$,  pointing out that the
system is mechanically unstable, as previously  discussed 
phenomenologically.\cite{Boyer:1985}

To show the main features of our analysis, we focus 
initially on the specific case of solid Argon 
and we generalize later these results for other rare-gas solid systems
(Ne, Kr, Xe), resulting in $T_S$  in fair agreement 
with the experimental values of the melting temperatures $T_m$. 

The paper is structured as follows. 
In Sec. \ref{s_model} we present our model for generic classical solids, recalling the SCPT 
for fixed volume and discussing  its extension for the fixed pressure case. 
The results for the specific case of solid Argon are shown in Sec. \ref{s_results}, 
pointing out how it is possible to achieve two qualitative and quantitatively 
different mechanical instabilities by working at constant volume or at constant pressure.
We also discuss there the singular behavior of the thermodynamical properties of the system. 
In Sec. \ref{s_phaseDiagram} we explain the phase diagram shown in Fig. \ref{f-phd}.
In the last Section \ref{s_concl} we drawn our conclusions.

\section{The model}
\label{s_model}

We consider a classical solid formed by $N$ particles
with mass $m$ and interacting via an isotropic
$U({\bf r})=U(r)$ pair  potential:
\begin{equation}
\label{eq:H}
H =\sum_i \frac{{\left| {\bf p}_i  \right|}^2}{2m}
+  \frac{1}{2} \sum_{i \neq j}
U\left( {\bf r}_i - {\bf r}_j  \right) \, ,
\end{equation}
where $i=1,\ldots,N$ and where  ${\bf p}_i$ and ${\bf r}_i$ are
the momentums and the positions of the particles.
The classical partition function of the system reads:
\begin{equation}
\label{eq:Z}
Z(T,V) =  \int \!\!\!
\int \prod_{i}   \frac{d {\bf r}_i  d {\bf p}_i}{h^3} \;\;
\mbox{e}^{-H / k_B T} \, ,
\end{equation}
which is a function of the total volume of the system $V$.
The statistical average for a generic physical quantity $O$ is given
by:
\begin{equation}
\label{eq:average}
\left< O \right> =  \frac{1}{Z}
\int \!\!\! \int \prod_i   \frac{d {\bf r}_i  d {\bf p}_i}{h^3}
O({\bf r}_i,{\bf p}_i)
\;\;  \mbox{e}^{-H / k_B T} \, .
\end{equation}
For sake of clarity, we consider here a monoatomic crystal with cubic symmetry.
At zero temperature the crystal is frozen and, in absence of any defect and 
far from boundary surfaces, the particles are fixed at their lattice positions
${\bf r}_i = {\bf R}_{i}$ with  $a=\left| {\bf R}_{i}-{\bf R}_{j}\right|$
 when $(i,j)$ are neighboring atoms. 
Because we work at fixed particle number, we have the simple relation 
$V \propto N a^3$.

At low temperature, the thermal fluctuations can be described  in the 
harmonic approximation where we expand the interatomic potential for small
lattice displacement of the particles ${\bf u}_i={\bf r}_i-{\bf R}_i$
around their average position ${\bf R}_i$.
By assuming a mean field approximation (the Einstein model), 
we can write:
\begin{equation}
\label{eq:H_Debye}
H_{\rm harm} =\sum_i \frac{{\left| {\bf p}_i  \right|}^2}{2m}
+  \frac{1}{2} \sum_{i \neq j}
U\left( R_{ij} \right)
+
\frac{1}{2} \sum_i  {\bf u}_i \cdot \hat{{\bf k}} \cdot  {\bf u}_i \, ,
\end{equation}
where $R_{ij}$ is a short notation for $R_{ij}=|{\bf R}_i-{\bf R}_j|$ and
$\hat{\bf k}$ is the elastic tensor responsible for the restoring force on
each particle.
In the cubic symmetry the tensor $\hat{\bf k}$ is diagonal
and isotropic in the three axial directions $\alpha=x, y, z$,
so that $k_{\alpha\beta}=k\delta_{\alpha\beta}$. 
We thus obtain the well-known result for the  mean thermal fluctuation
$\langle u^2 \rangle = 3 k_B T / k = 3 k_B T / m \omega_0^2$
where the Einstein frequency is $\omega_0=\sqrt{k/m}$.
As we raise the temperature to approach the melting point and 
beyond it (the overheated regime), anharmonic effects are expected 
to be relevant.
We can take them into account by using a variational method, the self 
consistent phonon theory, which we discuss in the next paragraph.

\subsection{The SCPT at fixed volume} 

We recall the standard scheme of the variational methods.
A (quadratic) trial Hamiltonian $H_{\rm v}$ containing variational parameters
is introduced. The corresponding partition function reads:
\begin{equation}
\label{eq:Z_T}
Z_{\rm v}(T,V) = 
\int \!\!\! \int \prod_i  \frac{d {\bf r}_i  d {\bf p}_i}{h^3} \;\;
\mbox{e}^{-H_{\rm v} / k_B T} \, .
\end{equation}
Then the Gibbs-Bogoliubov inequality gives as upper limit of $Z$:
\begin{equation}
\label{eq:Z_1}
Z  = Z_{\rm v}  {\langle \mbox{e}^{-(H-H_{\rm v})/k_B T} \rangle}_{\rm v}
\leq  Z_{\rm v} \; \mbox{e}^{- {\langle  (H-H_{\rm v})/k_B T \rangle}_{\rm v} } \, .
\end{equation}
where ${\langle \dots \rangle}_{\rm v}$ denotes the average as in Eq. (\ref{eq:average})
on the trial Hamiltonian $H_{\rm v}$. 
From Eq. (\ref{eq:Z_1}), we have the following inequality for the Helmholtz free 
energy $F(T,V)$:
\begin{equation}
\label{eq:F_v}
F = -k_B T \ln Z  \, \leq \,
-k_B T \ln Z_{\rm v} + {\langle  H-H_{\rm v} \rangle}_{\rm v} = F_{\rm v} \, ,
\end{equation}
where the r.h.s is the variational free energy $F_{\rm v}(V,T)$.
Then the variational parameters are determined
by minimization of $F_{\rm v}$.

We now apply the above described variational approach
within the context of the Self Consistent Phonon Theory (SCPT)
assuming a harmonic local model for any temperature.
We write
\begin{equation}
\label{eq:H_T}
H_{\rm v} =
 \frac{1}{2} \sum_{i \neq j}  U\left( {\bf R}_{ij} \right)
+ \sum_i \left( \frac{ {\left| {\bf p}_i \right|}^2 }{2m}
+  \frac{k_{\rm v}}{2}   {\left| {\bf u}_i \right|}^2 \right) \, ,
\end{equation}
where the local force constant $k_{\rm v}$
is the variational parameter.
It describes an {\em effective} elastic force
with frequency $\omega_{\rm v} =\sqrt{k_{\rm v}/m}$. 
We use now the hamiltonian $H_{\rm v}$, Eq. (\ref{eq:H_T}), to calculate 
the r.h.s. in Eq. (\ref{eq:F_v}), the variational free energy $F_{\rm v}(T,V)$  
whose the first term reads:
\begin{equation}
\label{eq:F_t}
-k_B T \ln Z_{\rm v} =
\frac{1}{2}  \! \sum_{i \neq j}  \! U\left({\bf R}_{ij} \right)
- 3 N k_B T \ln \left( \frac{k_B T}{\hbar \omega_{\rm v}} \right) \, ,
\end{equation}
By using the equipartition theorem for the quadratic term, we have also
directly the average of the exact Hamiltonian over the trial Hamiltonian:
\begin{equation}
\label{eq:H_m}
{\langle  H \rangle}_{\rm v} =
\frac{1}{2}  \sum_{i \neq j}
{\left<
U(|{\bf r}_i - {\bf r}_j|)
\right>}_{\rm v}
+
\frac{3}{2} N k_B T  \, ,
\end{equation}
and also the average of the trial  Hamiltonian:
\begin{equation}
\label{eq:H_T_m}
{\langle  H_{\rm v} \rangle}_{\rm v} =
\frac{1}{2}  \! \sum_{i \neq j}  \! U\left( {\bf R}_{ij} \right)
+
3 N k_B T \, .
\end{equation}
Summing up Eqs. (\ref{eq:F_t}), (\ref{eq:H_m}) and (\ref{eq:H_T_m}),
we finally get the variational free energy per particle:
\begin{equation}
\label{eq:F_v_1}
\frac{F_{\rm v}}{N}  =
- 3 k_B T
\ln \left( \frac{k_B T}{\hbar \omega_{\rm v}} \right)
+
\frac{1}{2N} \sum_{i \neq j} {\left<
U({\bf r}_i - {\bf r}_j)
\right>}_{\rm v}
-
\frac{3}{2} k_B T \, .
\end{equation}
It is convenient to introduce the smeared potential $\tilde{U}$ defined as:
\begin{eqnarray}
\tilde{U} \left( {\bf R}_{ij},u^2_{\rm v} \right)
&=&
{\left<  U({\bf r}_i - {\bf r}_j)   \right>}_{\rm v}
\nonumber \\
&=&
\int \!\!\! \frac{d{\bf k}}{ {\left(2 \pi \right)}^{3} }
U\left( {\bf k}  \right)
e^{i {\bf R}_{ij}\cdot {\bf k}}
 {\left<
\mbox{e}^{i \left( {\bf u}_i - {\bf u}_j \right) \cdot{\bf k}}  \right>}_{\rm v}
\nonumber \\
&=&
\int \!\!\! d{\bf r} \, U\left( {\bf r}  + {\bf R}_{ij}  \right)
\frac{ \, e^{- r^2/(4/3) u^2_{\rm v}}}
{ {\left(\frac{4}{3} \pi u^2_{\rm v} \right)}^{3/2}} \, , \label{eq:U_sm}
\end{eqnarray}
where $u^2_{\rm v}=\langle u^2 \rangle_{\rm v} $ is 
the thermal fluctuation in the SCPT:
\begin{equation}
\label{eq:u2_t}
u^2_{\rm v} =
 \frac{3 k_B T}{k_{\rm v}} \, .
\end{equation}
Eqs. (\ref{eq:F_v_1})-(\ref{eq:u2_t}) define at this stage
the explicit form of the variational free energy $F_{\rm v}$
which has to be minimized with respect to $k_{\rm v}$. 
Assuming isotropy along the $x$, $y$, $z$ directions, 
from the condition $d F_{\rm v} / d k_{\rm v} = 0$ 
we get:
\begin{eqnarray}
k_{\rm v}
&=&  -
\frac{1}{N}\sum_{i \neq j}
\int \!\!\! \frac{d{\bf k}}{ {\left(2 \pi \right)}^{3} }
U\left( {\bf k}  \right) \frac{|{\bf k}|^2}{3}
\mbox{e}^{i {\bf R}_{ij}\cdot {\bf k}}
{\left< \mbox{e}^{i \left( {\bf u}_i - {\bf u}_j \right)\cdot {\bf k}}
\right>}_{\rm v}
\nonumber \\
&=&  
\frac{1}{3N}
\sum_{i \neq j} \sum_{\alpha=x,y,z}
\frac{\partial^2 \tilde{U}({\bf R}_{ij},u^2_{\rm v})}{\partial R^2_{ij,\alpha}} \, ,
\label{eq:k_T}
\end{eqnarray}
which must be solved self consistently since the smeared potential 
$\tilde{U}$ depends implicitly on $k_{\rm v}$ via
Eqs. (\ref{eq:U_sm}), (\ref{eq:u2_t}).
In the case of a fully isotropic potential $U({\bf R}_{ij})=U(R_{ij})$, 
and reminding Eq. (\ref{eq:u2_t}), we get thus
the compact
self-consistent solution for $u^2_{\rm v}$:
\begin{equation}
\label{eq:selfcon}
\frac{3k_BT}{u^2_{\rm v}} 
 =  \frac{1}{3N}
\sum_{i \neq j}
{\left. \left(
\frac{d^2 \tilde{U}(x,u^2_{\rm v})}{dx^2} + \frac{2}{x} \frac{d \tilde{U}(x,u^2_{\rm v})}{dx}
\right)\right|}_{x= R_{ij}} .
\end{equation}
The SCPT allow us to evaluate in a self-consistent way
the mean thermal fluctuations $u^2_{\rm v}$
(and hence the effective elastic constant $k_{\rm v}$)
for given temperature.
In the literature,\cite{Choquard:1967,Gillis-1:1968,Gillis-2:1968,Kugler:1969,Klein:1970,Gillis:1972,
Matsubara:1977,Pietronero:1979,Hasegawa:1980,Jayanthi-1:1985,Jayanthi-2:1985,Gong:1987,Moleko:1983} 
such an approach has been employed in the analysis of lattice 
mechanical stability against anharmonic  fluctuations as temperature raises  
 at fixed volume.\cite{note}
The temperature of instability was thus identified with the maximum 
temperature  $T^*$ above which a self-consistent solution for
$u^2_{\rm v}$ disappears.

\subsection{Extension at fixed pressure}

After the minimization, the SCPT provides us an approximated expression 
for the Helmholtz free energy  $F(T,V) \simeq F_{\rm v}(T,V)$ as function of the volume $V$.
In order to describe how the volume per particle $V/N$ evolves with temperature and
pressure, we introduce in the usual way the Gibbs free energy $G$ defined as 
$G(T,P) = F_{\rm v}[T,V(T,P)] + P \, V(T,P)$, where now the volume $V=V(T,P)$ 
has to be considered itself as a function of temperature and pressure,
and it is given by the relation ${\left( \partial G/\partial P \right)}_T= V$ 
or, equivalently, by inverting the relation
\begin{equation}
\label{eq:pressure}
{\left. \frac{\partial F_{\rm v}(T,V)}{\partial V}\right|}_T = - P \, ,
\end{equation}
which gives an implicit definition of the volume $V=V(T,P)$.

\section{Results}
\label{s_results}

\begin{figure}[t]
\begin{center}
\includegraphics[scale=0.3,angle=270.]{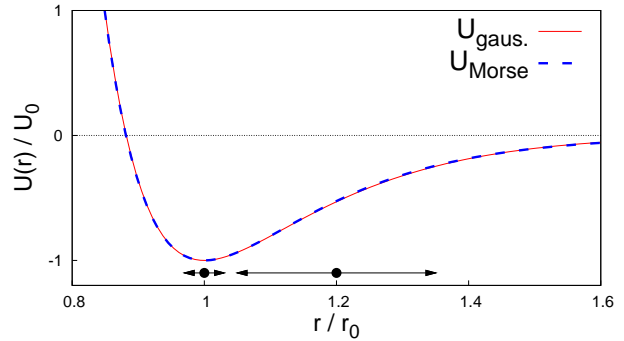}
\caption{Comparison between the Morse and the Gaussian-like potential for the case
of  Argon. 
Model parameters are here $r_0 = 3.76$ \AA, $U_0=146.8$ K and $\gamma=1.55$ \AA$^{-1}$ 
for the Morse potential, and $r_0 = 3.76$ \AA, $U_0=146.8$, $\beta_+= 0.45$ and 
$\beta_-=0.16$ 
for the Gaussian-like model, $\beta_+,\beta_-$ obtained by the best fitting 
of the Morse potential.
The filled circles and horizontal double arrows 
represent the lattice parameter and the corresponding
lattice fluctuations at low temperature
($T \simeq 1$ K) where $a \approx 0.98 r_0$
and at the instability temperature $T\approx T_S=227$ K where, 
for $P=0$, we have $a \approx 1.2 r_0$ due to thermal expansion.}
\label{fig:potential}
\end{center}
\end{figure}

In order to keep the calculations at the most analytical level, 
we assume in the following a particularly suitable 
form for the potential $U\left( {\bf r} \right)$ which reproduces 
the standard molecular potentials in the whole effective 
range for distance experienced by the particles at zero temperature as well as
in the vicinity of the instability temperature $T_S$. 
We consider in particular a potential given by the linear
combination of two Gaussian representing a long range attractive 
tail and a short-range repulsion:\cite{Hasegawa:1980}
\begin{equation}
\label{eq:potential}
U(r) =  \frac{U_0}{\beta_{+} - \beta_{-}}
\left(
\beta_{-} \mbox{e}^{- \beta_{+} (r^2- r_0^2)}
-
\beta_{+} \mbox{e}^{- \beta_{-} (r^2- r_0^2)}
\right) \, ,
\end{equation}
where $r=|{\bf r}|$ and where $U_0$ is the potential minimum at the point $r=r_0$.
By a proper choice of the two parameters $\beta_{+},\beta_{-}$ it is thus possible
to reproduce the behavior of the standard molecular potentials as, for instance,
the Lennard-Jones or the Morse potential in the range which we are interested. 
We stress that the two parameters $\beta_{+},\beta_{-}$ are independent on temperature.  
An example is given in Fig. \ref{fig:potential}, where we compare the Gaussian-like 
potential of Eq. (\ref{eq:potential}) with the Morse potential  describing 
Argon.\cite{Moleko:1983} 

In Fig. \ref{fig:potential}  the lattice parameter $a$ (black dots)and the
range of the corresponding lattice fluctuations (double arrows) 
are also shown for  $T \approx 0$ and for $T = T_S(P=0) \simeq 227$ K 
as representative cases of low and high temperature respectively.
As we see, the agreement between the  Morse potential for Argon and our 
Gaussian-like potential is practically perfect in all the physical range of $r$ spanned
at both low and high temperature.
This agreement is particularly important since it assures
that the results of the variational method for two potentials 
are indistinguishable.

For the Gaussian-like form of Eq. (\ref{eq:potential}), the smeared potential reads:
\begin{eqnarray}
\label{eq:U_sm_bulk}
&&\tilde{U} \left(r \right)
=
\frac{U_0}{\beta_+-\beta_-}
\sum_{\sigma=\pm}  \frac{ \sigma  \beta_{-\sigma} }{{\left( 1 + \frac{4}{3} \beta_{\sigma} u^2_{\rm v} \right)}^{3/2}}
\nonumber\\
&&
\times
\exp\left[ - \beta_{\sigma}
\left( \frac{ r^2 }{ 1 + \frac{4}{3} \beta_{\sigma} u^2_{\rm v} } - r_0^2 \right) \right] \, .
\end{eqnarray}
In the zero temperature limit ($u^2_{\rm v} = 0$) the smeared potential 
reduces to the bare potential $\tilde{U}=U$ and the elastic constant Eq. (\ref{eq:k_T}) 
to the bare elastic constant $k_{\rm v}=k_0$.
The zero temperature/zero pressure  nearest neighbors distance $a_0=a(T=0,P=0)$
is found  by minimizing the classical ground  state energy 
$E_0 = (1/2) \sum_{i \neq j} U\left(R_{ij} \right)$.
For the Argon parameters considered in this section we get
$k_0=350.5 \, U_0/r_0^3$ and $a_0=0.98 \, r_0$ which is in good agreement
with the experimental value $a_0^{\rm exp.}=0.99 \, r_0$.\cite{Peterson:1966} 
Note that the value $a_0=0.98 \, r_0$ is slightly different from the potential 
minimum $r_0$ due to the contribution  of the second nearest neighbors atoms in the total energy.
By increasing the temperature, both the thermal fluctuations $u_{\rm v}^2$ and
the lattice parameter $a$ will have a non trivial dependence on temperature.

\subsection{Constant volume}

\begin{figure}[h]
\begin{center}
\includegraphics[scale=0.3,angle=270.]{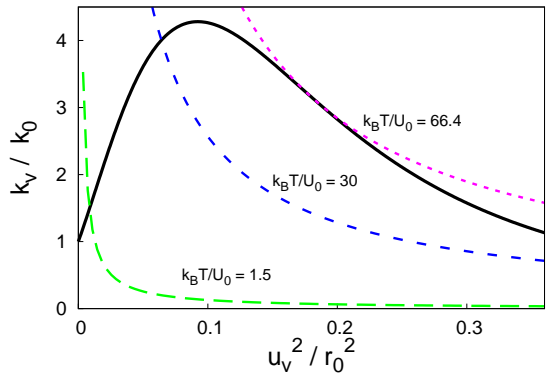}
\caption{
Graphical solution of the self-consistent equation Eq. (\ref{eq:selfcon})
for fixed lattice parameter $a=a_0$ and for the Argon parameters.
The dashed lines correspond to the
l.h.s. of Eq. (\ref{eq:selfcon})  for three representative temperatures
while the solid line represents 
the r.h.s. of Eq. (\ref{eq:selfcon}) which does not depend
parametrically on the temperature.
}
\label{fig:selfEquation}
\end{center}
\end{figure}

We discuss first the case where the lattice parameter
$a$ is assumed to be independent of $T$ and set to the value $a=a_0$, i.e. its value 
at zero temperature and zero pressure. 
In this case, the effective elastic constant $k_{\rm v}(T)$ and the lattice
fluctuations $u_{\rm v}^2(T)$ can be simply obtained as functions of the
temperature by the self-consistent solution of
Eqs. (\ref{eq:U_sm})-(\ref{eq:selfcon}), setting the $a=a_0$ in the lattice sum.
The graphical solution of Eq. (\ref{eq:selfcon}) for three representative
temperatures is shown in Fig. \ref{fig:selfEquation}.

\begin{figure}[h]
\begin{center}
\includegraphics[scale=0.3]{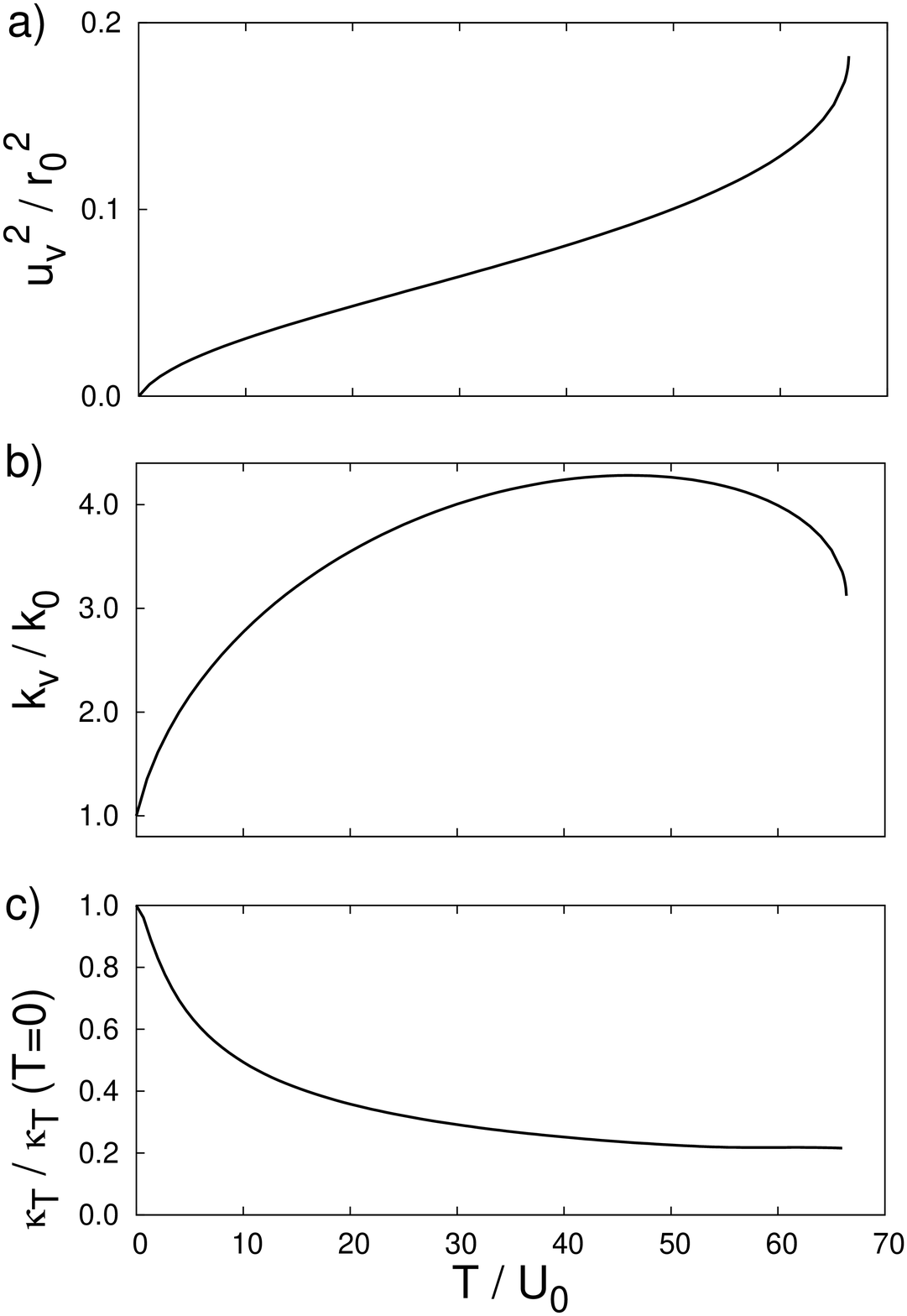}
\caption{
(a) Mean lattice fluctuations $u_{\rm v}^2(T)$, 
(b) effective elastic constant $k_{\rm v}(T)$, 
solutions of the self-consistent equation (\ref{eq:selfcon})
for fixed lattice spacing $a=a_0$ and for the Argon parameters. 
(c) The compressibility $\kappa_T$ as function
of temperature.}
\label{fig:results_a0_1}
\end{center}
\end{figure}

The physical quantity $u_{\rm v}^2$ (and hence $k_{\rm v}$)
is obtained from the lowest-$u$ intersection of the l.h.s function and of the
r.h.s. function which is an implicit function of $u_{\rm v}^2$.
In the zero temperature limit this solution corresponds to the harmonic 
limit $u_{\rm v}^2(T) \simeq 3k_BT/k_0$.  
On the contrary, the second solution, which is 
not shown in Fig. \ref{fig:selfEquation}, starts at $T \simeq 0K$ from 
values much higher than the nearest neighbors distance $a$ and 
then it decreases with temperature. 
It corresponds  to a local maximum which does not minimize the
free energy and it can be disregarded.
The behaviors of $u_{\rm v}(T)$  and $k_{\rm v}^2(T)$  as functions
of the temperature are plotted in Fig. \ref{fig:results_a0_1}(a) and in Fig. \ref{fig:results_a0_1}(b).

For a temperature independent volume, above an upper temperature $T^*(a_0)$ 
no solution is found (see Fig.~\ref{fig:selfEquation}).
The maximum temperature found in this way at $a=a_0$ for the Argon
potential parameters is $T^*=66.4 \, U_0 \approx 10000$ K,
much larger than the experimental melting temperature of the Argon 
$T_m  = 0.56 \, U_0\approx 82$ K.
Note that the temperature $T^*$ is so high
that the linear behavior of $u^2_v(T) \propto T$
(valid for an harmonic crystal) is
confined to a very small temperature range $T \ll  T^*$ (Fig. \ref{fig:results_a0_1}(a)).

We would like to stress that, as remarked in Ref. \onlinecite{Pietronero:1979}, 
the breakdown of the SCPT solution at $T^*$ is not signalized by any
precursor singular behavior in $u^2_{\rm v}$ nor in
$k_{\rm v}$.
On this regards it is  worth to analyze explicitly even the behavior of the
compressibility $\kappa_T=-(\partial P/\partial V)_T/V$
as function of the temperature close to the instability
temperature $T^*$ where the solution of Eq. (\ref{eq:selfcon})
disappears.
The SCPT solution for the compressibility $\kappa_T$ is thus reported in 
Fig. \ref{fig:results_a0_1}(c), showing that $\kappa_T$
is well-behaved as the instability temperature at fixed volume $T^*$
is approached.
As we are going to see, the behavior of $\kappa_T$ is qualitatively different when
the thermal lattice expansion as a function of the temperature at fixed 
pressure is considered.

\subsection{Constant pressure}

Before discussing explicitly a system at constant pressure, 
let us consider first the interatomic distance $a$ as an external and 
tunable parameter in order to gain some useful preliminary insight. 

\begin{figure}[t]
\begin{center}
\includegraphics[scale=0.3,angle=270.]{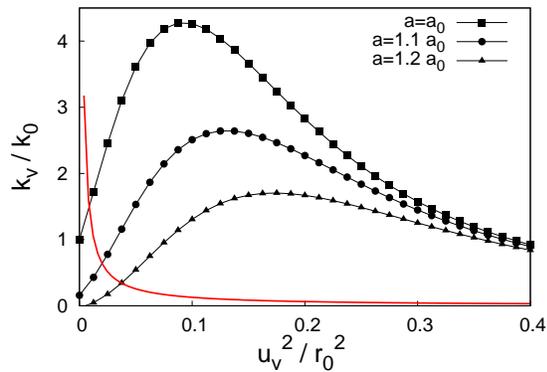}
\caption{
Graphical solution of the self consistent equation Eq. (\ref{eq:selfcon})
at given temperature $T=1.5 \, U_0$ and at different lattice spacing  $a\ge a_0$.
The solid line and the dotted lines  represent, respectively, 
the l.h.s. and the r.h.s. function of Eq. (\ref{eq:selfcon}) for the Argon parameters.
}
\label{fig:kt_function_ann}
\end{center}
\end{figure}

In Fig. \ref{fig:kt_function_ann} we show the  behavior of the graphical 
solution of Eq. (\ref{eq:selfcon}) {\em for fixed temperature} as varying $a$.

We observe that the global effect of increasing the lattice spacing is to 
reduce the effective elastic constant $k_{\rm v}$ and to increase thus the 
mean lattice fluctuations $u_{\rm v}^2$. 
Thus, according to the Lindemann criterion, we can argue that the breakdown of 
the solid phase can occur at temperatures lower than the one corresponding to 
the constant volume case.

\begin{figure}[h]
\begin{center}
\includegraphics[scale=0.3,angle=270.]{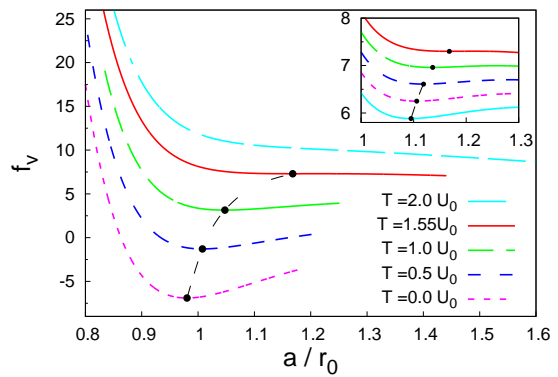}
\caption{
Dimensionless free energy $f_{\rm v}=F_{\rm v}/NU_0$ as a function of $a$
in the Argon case for different temperatures.
The filled symbols represent the minima of $F_{\rm v}/NU_0$ and the
dot-dashed line shows the their evolution with T.
Inset: a close-up for $T\approx T_S=1.55 \, U_0$ $(T=1.35,1.40,1.45,1.50,1.55 \, U_0)$.
For sake of visibility, in the main panel and in the inset each curve of the dimensional 
free energy  has been shifted by a factor $F_{\rm v}/NU_0 \rightarrow
F_{\rm v}/NU_0+3k_BT/2U_0$, which does not affect however the
determination of $a$ at the minima.
}
\label{fig:variationalF}
\end{center}
\end{figure}

In a more compelling way, the physical value of the lattice parameter $a(T)$
(or the volume per particle $V(T)/N $) is obtained  for a given pressure $P$ and temperature $T$ 
by inverting the relation $\partial F_{\rm V}(T,V)/\partial V|_T=-P$, Eq. (\ref{eq:pressure}).
In Fig. \ref{fig:variationalF} we show as an example the case of zero pressure 
$P=0$ which simply corresponds to the minimization
of the free energy $F_{\rm v}$ with respect to the volume $V$ or,
equivalently, to $a$.
The evolution of the minimum of $F_{\rm v}$ at different temperatures
shows the thermal expansion of the system at constant pressure $P=0$.

\begin{figure}[t]
\begin{center}
\includegraphics[scale=0.3]{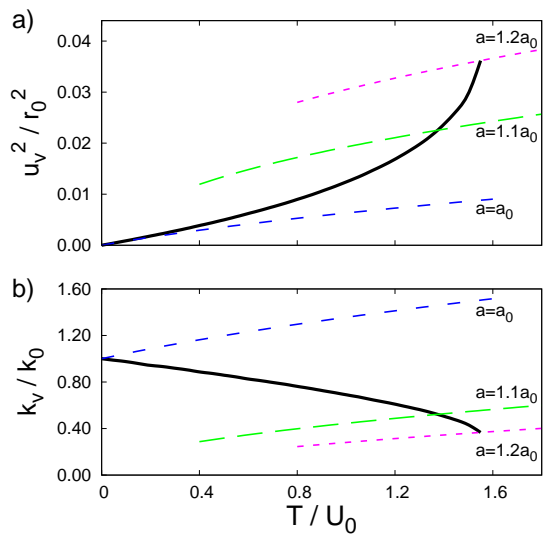}
\caption{
(a) mean lattice fluctuations $u_{\rm v}^2$ and (b) 
effective elastic constant $k_{\rm v}$  
as functions of temperature at constant pressure $P=0$,
obtained from the minimization of $F_{\rm v}$ with respect to
$a(T)$ as illustrated in Fig. \ref{fig:variationalF}. 
The dashed lines correspond to the solution of the self-consistent
equations for fixed lattice spacing $a=1.0, 1.1, 1.2 a_0$.
}
\label{fig:kt_u2_aT}
\end{center}
\end{figure}

The corresponding results for the elastic constant $k_{\rm v}$ and for the thermal 
fluctuations $u^2_{\rm v}$ are reported in Fig. \ref{fig:kt_u2_aT}(a) and in Fig. \ref{fig:kt_u2_aT}(b).
For comparison in Fig. \ref{fig:kt_u2_aT} it is also shown the behavior
of $u^2_{\rm v}$ and $k_{\rm v}$  obtained by the self-consistent solution 
of Eq.(\ref{eq:selfcon}) as functions of the temperature but assuming three 
representative fixed lattice spacing $a$.
Note that the physical behavior of the elastic constant $k_{\rm v}$ as a function 
of the temperature is quite different in the two cases, with a {\em decrease} of
$k_{\rm v}$ as a function of $T$ at fixed pressure whereas 
a hardening of $k_{\rm v}$  is predicted when working at fixed volume

Our calculated data in Fig. \ref{fig:kt_u2_aT} (case $P=0$)
extend up to the spinodal temperature $T_S\approx 1.55 U_0$,
above which the minimum of $F_{\rm v}$ as a function of $a$ disappears.
It is worth noting that such temperature $T_S\approx 1.55 \, U_0$
is much lower than   $T^* \approx 66.4 \, U_0$ related to existence of the self-consistent solution 
for the lattice fluctuation, and it is of the same order of magnitude 
of the experimental melting temperature $T_{\rm m}\approx 0.5 \, U_0$.

\begin{figure}[t]
\begin{center}
\includegraphics[scale=0.3]{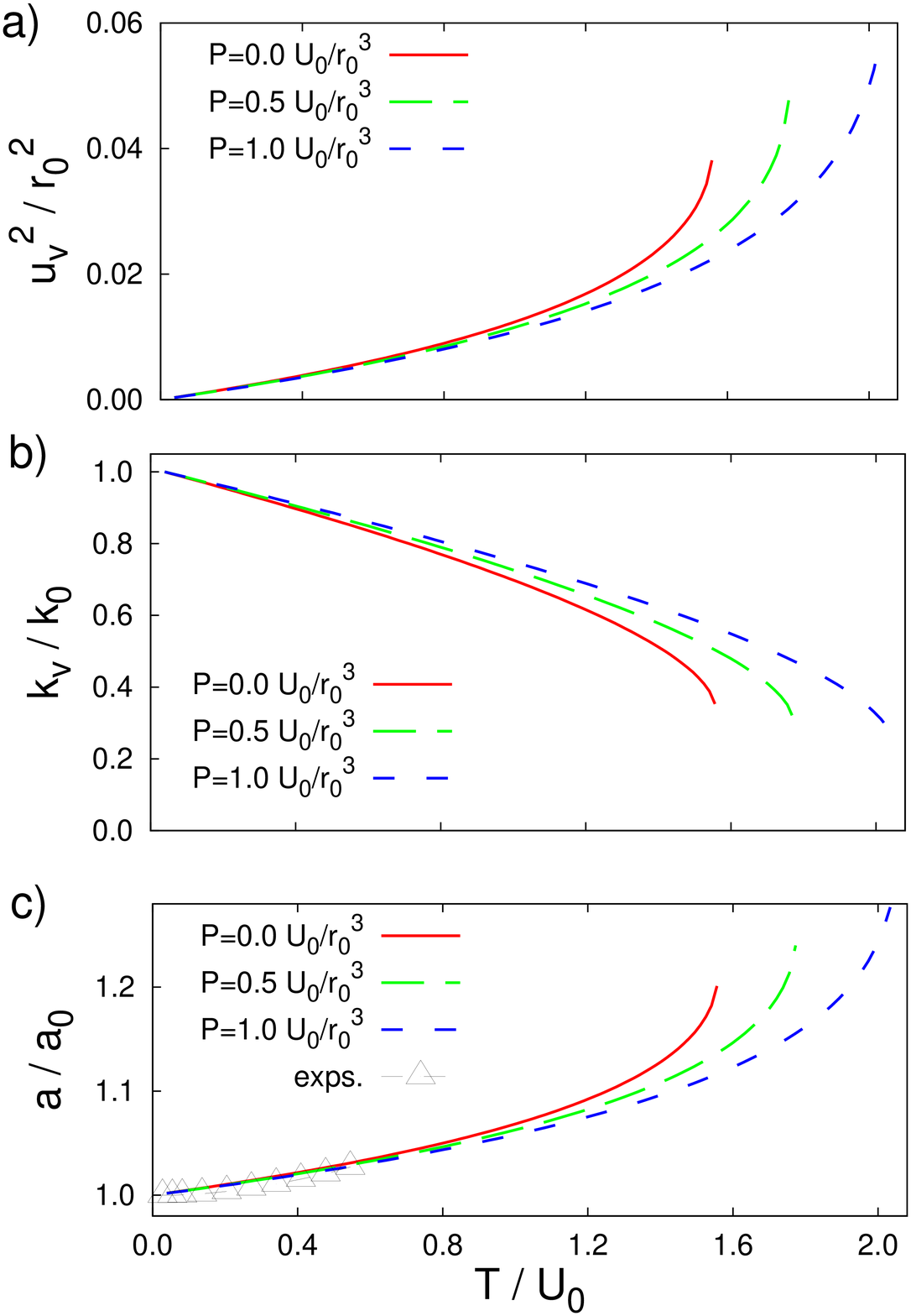}
\caption{(a) Mean lattice fluctuations $u_{\rm v}^2$, 
(b) effective elastic constant $k_{\rm v}$ and (c)
lattice parameter $a$ as functions of $T$ for $P=0.0, 0.5, 1.0 \, U_0/r_0^3$.
The corresponding spinodal temperatures are 
$T_S =1.55, 1.80, 2.05 \, U_0 $. 
The triangles in (c) correspond to experimental data 
after Ref. \onlinecite{Peterson:1966}.
}
\label{fig:P_kt_u2_aT}
\end{center}
\end{figure}

Similar results are obtained for fixed finite pressure $P \neq 0$. 
As mentioned in Sec. \ref{s_model}, the thermal lattice expansion
can be traced by minimizing the Gibbs free energy or, equivalently,
by inverting Eq. (\ref{eq:pressure}) to obtain $a$ as a function
of $T$ and $P$.
The behavior of the average fluctuation, the effective 
elastic constant  and the lattice constant are reported in Fig. \ref{fig:P_kt_u2_aT}.
Available experimental data of $a(T)$ in Argon, in an experimental setup at saturated 
conditions corresponding to $P \approx 10^{-3} U_0/r_0^3\approx 0$,\cite{Peterson:1966}
are also reported in Fig.\ref{fig:P_kt_u2_aT}(c), 
showing an agreement between our calculations  and the experimental data.
We observe here that increasing the pressure leads to  a weaker increase 
of $u_{\rm v}^2$ (weaker decrease of $k_{\rm v}^2$) as a function of $T$. 
In other words, the finite pressure leads to an increase of the spinodal temperature $T_S(P)$ 
at which the solid phase disappears as a metastable solution. 
We also note that, in similar way as in the analysis at constant volume,
neither $u_{\rm v}^2$, $k_{\rm v}$, nor the lattice constant $a$
show any singular behavior at $T_S(P)$.

\begin{figure}[t]
\begin{center}
\includegraphics[scale=0.3,angle=270.]{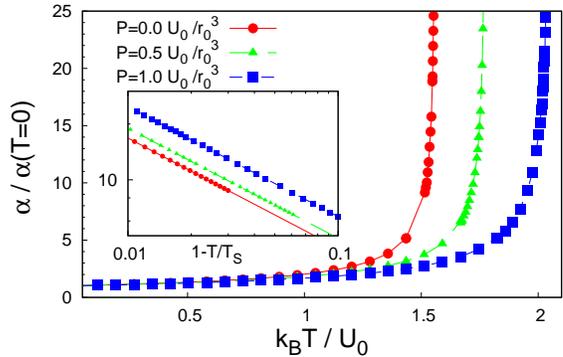}
\caption{
Temperature behavior  of the thermal expansion coefficient $\alpha/\alpha(T\!=\!0)$
for three representative pressures $P=0.0\,, 0.5\,, 1.0 \, U_0/ r_0^3$.
Inset: logarithmic plot of $\alpha/\alpha(T\!=\!0)$ vs. the reduced
variable  $t=|1-T/T_S|$, showing the power law behavior with
$\alpha \approx t^{-\zeta}$.}
\label{fig:thermal_dilatation}
\end{center}
\end{figure}

In the present case at constant pressure, however, an additional tool of investigation is provided
by the thermal lattice expansion  coefficient $\alpha$ defined as:
\begin{equation}
\label{eqn:alpha}
\alpha =  \frac{1}{V} {\left. \frac{\partial V}{\partial T} \right|}_P.
\end{equation}
The temperature dependence of $\alpha$ at fixed temperature is reported in
Fig. \ref{fig:thermal_dilatation}, showing the singular behavior at $T_S$. 
In the inset we plot the thermal expansion 
coefficient $\alpha$ as a function of the reduced  variable $t=|1-T/T_S|$ 
in a log-log scale, showing a power-law divergence $\alpha \approx t^{-\zeta}$,
where $\zeta \simeq 0.6$ was obtained by a fitting procedure.

\begin{figure}[b]
\begin{center}
\includegraphics[scale=0.3,angle=270.]{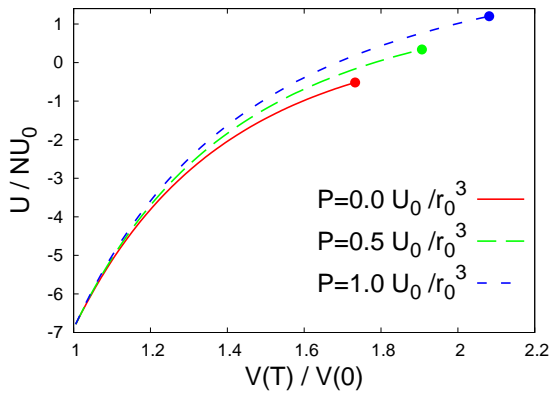}
\caption{Internal energy $U$ as a function
of the volume $V(T)$ varying the temperature for different
pressures $P=0.0,\, 0.5,\, 1.0 \, U_0/r_0^3$.}
\label{fig:dU-dV}
\end{center}
\end{figure}

At the spinodal temperature, the divergence of the lattice expansion 
coefficient $\alpha$ coincides with a singular behavior of the 
compressibility $\kappa_T$ as it could be guessed by the general 
thermodynamical relation:\cite{Huang:2002}
\begin{equation}
\label{eqn:therm_relation}
\alpha^2 = \frac{C_P-C_V}{V T} \kappa_T  \, ,
\end{equation}
where $C_p$ and $C_V$ are respectively the specific heat at
constant pressure and constant volume. 
Note that a general condition for the lattice  stability is the positiveness of the 
two specific heats  ordered as $C_P>C_V>0$.\cite{Landau:2002}
Using thermodynamical relations, 
we can rewrite Eq. (\ref{eqn:therm_relation}) in the more convenient form
\begin{equation}
\alpha^2
=
\frac{1}{T}\left( P+\frac{\partial U}{\partial V} \right)\kappa_T^2 \, ,
\label{eqn:therm_relation_2}
\end{equation}
where the internal energy per particle is defined in the Eq. (\ref{eq:H_m}), 
$U=\langle  H \rangle_{\rm v}$.
From Eq. (\ref{eqn:therm_relation_2}), 
the divergence of $\alpha$ at the spinodal temperature $T_S$
implies the divergence of $\kappa_T$ provided the quantity $\partial U/\partial V$ 
has a finite and regular behavior as $T$ approaches $T_S$.
As can be seen from Fig. \ref{fig:dU-dV}, both $U$ and its derivative 
$\partial U / \partial V$  have a regular behavior for any value of temperature 
and pressure, pointing out that $\alpha$ is proportional to $\kappa_T$ 
at the spinodal temperature $T_S$. 

\section{Phase Diagram}
\label{s_phaseDiagram}

In the previous Sections, we have identified two different mechanisms 
for the breakdown of the solid state phase within the SCPT model, 
working respectively at fixed volume and at fixed (moderate) pressure. 
In the first case the solid phase instability is pointed out by the breakdown of 
the self-consistent solution of the SCPT at a temperature $T^*$, 
with no other evident signature in the thermodynamical properties. 
On the other hand, a solid state phase is always defined
in the second case, which however disappears as a metastable minimum
at a spinodal temperature $T_S$, where the thermal expansion
coefficient $\alpha$ and the compressibility $\kappa_T$ diverge.

\begin{figure}[b]
\begin{center}
\includegraphics[scale=0.3,angle=270.]{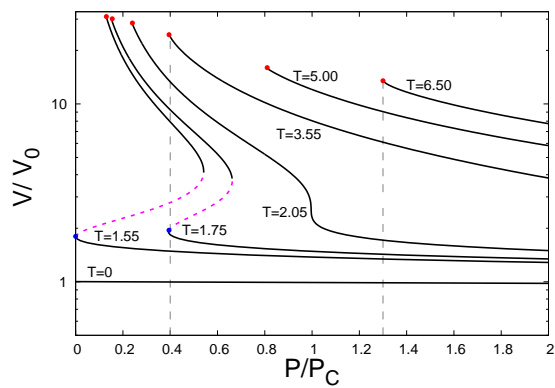}
\caption{(color online) 
Representative isothermal curves
of different regimes, for 
$T = 0, 1.55, 1.75, 2.05, 3.55, 5.00, 6.50 \, U_0$.
Solid line regions represent the case where the compressibility
$\kappa_T$ is positive, while dashed line regions
correspond to $\kappa_T<0$.
The volume is expressed in units of the
$V_0=Na_0^3$, and the pressure in units
of the critical pressure $P_c=1.02 \, U_0/r_0^3$, whose
meaning is discussed in the text.
}
\label{fig:V_vs_P}
\end{center}
\end{figure}

The competition between these two different behaviors
can be better clarified analyzing in more details 
some typical isothermal curves in a wide range of temperatures as shown 
in Fig. \ref{fig:V_vs_P}. 
We can thus distinguish between two different pressure ranges.

For high pressures  ($P > P_c$ with $P_c \approx 1.02 U_0/r_0^3$) we have only 
one volume solution for given pressure by increasing temperature. 
In this range  the isothermal curves are monotonic as function of pressure,
corresponding to a positive isothermal compressibility. 
This behavior holds true at high temperatures ($T\gtrsim 2 U_0$)  for any pressure.
Increasing the temperature at constant pressure leads to a rapid increase 
of the volume until the self-consistent
solution of the SCPT disappears at the temperature $T^*$.
At $P \simeq 1.3 P_c$ we have for instance $T^*=6.50\, U_0$ (Fig. \ref{fig:V_vs_P}).

Below the critical pressure $P_c$, 
 the isothermal curves are non-monotonic with three possible volume solutions for given pressure,
and an intermediate region $V_1(P) < V < V_2(P)$ of negative compressibility
$\kappa_T<0$ (dashed lines in Fig. \ref{fig:V_vs_P}).
In this case,  the physical more stable solution is the one with
smaller volume.\cite{note_PS}  
Increasing the temperature thus leads to a slight increase of $V$ until 
the first physical solution disappears at the spinodal temperature $T_S$ where
the $dP/dV=0$, and the compressibility $\kappa_T$ diverges 
(for $P\simeq 0.4$, we have $T_S \approx 1.75\, U_0$ with 
corresponding to a volume $V\approx 2 V_0$, see Fig. \ref{fig:V_vs_P}). 
At higher temperature, disregarding phase separated phases,
the only physical solution is associated with the large volume
cases (for $P\simeq 0.4$, we have  $V\approx 8 V_0$ for $T \gtrsim 1.75 U_0$).
This sudden jump of the volume at $T=T_S$ reflects thus the divergence of
the lattice expansion coefficient $\alpha$ at the spinodal temperature.

Increasing further the temperature, even this high volume solution
will break due to the breakdown of the self-consistent SCPT solution
at a high temperature $T^*$.
As a matter of facts, this kind of instability is driven by the
disappearing of the SCPT solution at $T^*$ (the same encountered when working 
at constant volume) 
in contrast to the compressibility divergence at $T_S$.

\begin{figure}[t]
\begin{center}
\includegraphics[scale=0.3,angle=270.]{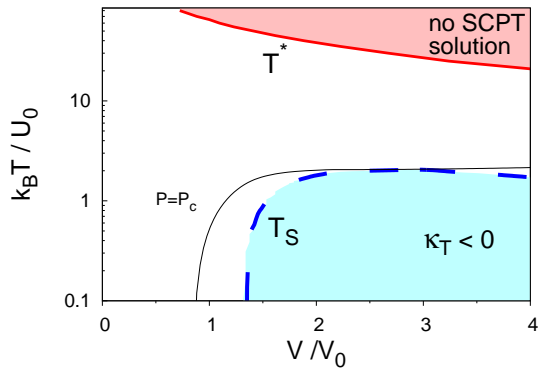}
\caption{(color online)
Phase diagram for the solid instabilities in the volume-temperature space.
The SCPT solution disappears for $T>T^*$.
The long dashed line represents the
spinodal temperature $T_S$ where the compressibility
is negative ($\kappa_T<0$ blue shadow region)
for $T<T_S$.
Also shown is the isobaric curve (tiny solid black line)
corresponding to $P=P_c$. 
}
\label{fig:PhD}
\end{center}
\end{figure}

We can summarize all the above discussion in a compact phase diagram, Fig. \ref{fig:PhD}, 
which reproduces on a quantitative level the schematic phase diagram sketched in Fig. \ref{f-phd}. 

\begin{figure}[t]
\includegraphics[scale=0.3,angle=270.]{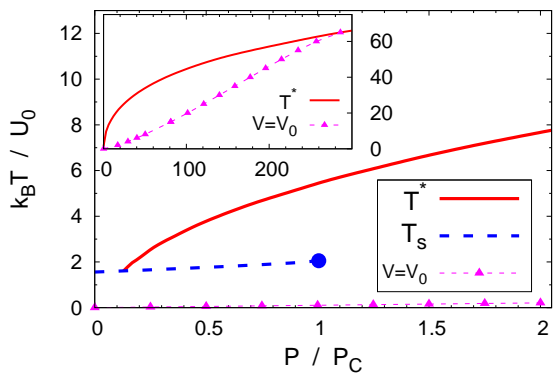}
\caption{Pressure dependence of $T_S$ and $T^*$
for the Argon case. Also shown as symbols
is the $P$ vs. $T$ dependence of the isochoric line for
$V=V_0$. Inset: same quantities in a large axes scale, in order to
show the intercept of the isochoric line with $T^*$ which
defined the solid phase instability at fixed volume $V=V_0$.}
\label{fig:Ti_line}
\end{figure}

The different kinds of instabilities of the solid phase
as a function of the pressure are depicted in 
Fig. \ref{fig:Ti_line}, where the dashed line represents
the spinodal temperature $T_S$ accompanied by
the divergence of the lattice expansion coefficient $\alpha$
and of the compressibility $\kappa_T$, while the solid line
marks the disappearing of the SCPT solution, as function of temperature, 
not reflected in any singular behavior of the thermodynamical properties.
Note that for $T<T_S$, $T^*$ is not defined.

The physical mechanism of the breakdown
of the solid phase is determined by the first instability 
encountered by increasing the temperature.
For $P<P_c$ this is given by the compressibility divergence at $T_S$.
For $P>P_c$ no thermodynamical instability is encountered
and the unique instability of the solid phase is associated
with $T^*$.

Fig.\ref{fig:Ti_line} permits also to clarify in simple terms 
why the mechanical instability at constant volume
occurs at much higher temperature $T^*$ and with different
phenomenology than at constant pressure.
Considering for simplicity a constant volume $a=a_0$ ($V=V_0$),
increasing the temperature is indeed associated
with an effective increase of $P$. 
Such increase is however negligible on the scale $P \approx P_c$, so that the first encountered
solid phase instability occurs at much higher temperatures,
as shown in the inset, and it is related to the SCPT breakdown
at $T^*$.

\section{Discussion and conclusions}
\label{s_concl}

In this paper we have revisited the self-consistent phonon theory (SCPT) 
to analyze the mechanism of the mechanical instability of an overheated 
metastable solid at finite pressure. 

Within this approach, we have predicted that two different instabilities 
are possible for the mechanical melting.
They occur at very different temperatures, depending whether working at fixed volume or pressure.
In the first case the instability of the solid phase occurs 
at rather high temperatures and it is associated with a breakdown of the SCPT 
solution at $T^*$, not reflected in any singular behavior of the thermodynamical
quantities. 
On the other hand, the mechanical instability approached at fixed pressure,
for pressures smaller than a critical pressure $P_c$,  can be properly interpreted 
in terms of a spinodal temperature $T_S$ at which the solid phase, always defined, 
disappears as a metastable minimum. 
In the latter case the mechanical instability is reflected in the divergence 
of the thermal compressibility  $\kappa_T$ as well as in the divergence 
of the lattice expansion coefficient $\alpha$ at $T=T_S$.  
Such kind of spinodal instability disappears for $P>P_c$, where only the instability 
at $T^*$ driven by the breakdown of the SCPT solution remains.
The value of $P_c$ is found to be of order $P_c \sim U_0/a_0^3$ where $U_0$ 
is the energy minimum of the pair potential and  $a_0$ the lattice spacing. 
For the weakly bound rare-gas solids we have $U_0$ of the order of few
meVs, corresponding thus to a critical $P_c$ of the order of some MPa.

For a quantitative comparison with real systems, in this paper we have mainly focused
on the specific case of an Argon classical solid
where intensive theoretical and experimental investigation
have been performed in literature (as example, 
see Ref. \onlinecite{Luo:2007} and references therein).

Similar analysis can be however carried out in a simple way 
as well for other  rare-gas solids (Xe, Ne, Kr). 
The results are  collected in Table \ref{tab:summary} where we report
the theoretical  $T^*$ and $T_S$ compared also with
the experimental melting temperature $T_m$.
As we can see,  $T_S$ is systematically much lower than $T^*$
and it is  of the same order of magnitude of the experimental 
melting temperature $T_m$.
Inclusion of higher order anharmonic terms\cite{Paskin:1982} and the development of models 
beyond the Einstein one\cite{Moleko:1983,Shukla:1998} 
(i.e. taking into account the full phonons dispersion) might further reduce $T_S$
towards the empirical range for overheating $T_{\rm max}\sim 1.5 \, T_m$.\cite{Luo-2:2003}

\begin{table}[hbtp]
\begin{tabular}{l c r r r r r}
\hline\hline
    & & $T_{m}$ (K) &     & $T^*$ (K) &     & $T_S$ (K)\\
\hline
Ne & &           25      & & 2450    & &    69  \\
Ar & &           83      & & 9747    & &   228  \\
Kr & &          116      & &18986    & &   317  \\
Xe & &          161      & &21662    & &   445  \\
\hline\hline
\end{tabular}
\caption{Experimental melting temperature $T_m$ of (Ne, Ar, Kr, Xe)
compared to the temperatures of the mechanical lattice instability as
obtained by the SCPT at fixed volume ($T^*$) with $a=a_0$  and 
at fixed pressure ($T_S$) with $P=0$ .
}
\label{tab:summary}
\end{table}

Such analysis points out that taking into account the lattice
expansion  is of fundamental importance not only for a proper quantitative
estimate of the instability temperature of the solid phase, 
but also for revealing the different instability mechanisms
associated or not associated with a singular behavior
of the thermodynamical properties.

\acknowledgments 
The authors acknowledges useful discussions with S. Ciuchi. 
Critical reading of the manuscript by M. Holzmann is appreciated.


\begin{thebibliography}{99}
\bibitem{note}
An exception is the work of Ref. \onlinecite{Moleko:1983}, where 
the authors took into account the thermal dilatation of the 
lattice by the experimental data but, for temperatures beyond the melting temperature
$T_m$, they assumed the lattice spacing frozen at the value $a=a(T_m)$.
\bibitem{note_PS}
A Maxwell construction, corresponding to a phase separation
between coexisting regions of stable solid phases with different
volumes is in principle possible in this framework. 
The analysis of such possibility goes however beyond the purposes
of the present work, which is aimed to underlying the different
breakdown mechanisms of the solid phase within the SCPT
operative at constant $V$ and at constant $P$.
\end{thebibliography}
\end{document}